\documentclass[12pt]{article}
\usepackage{amssymb}
\def\beq{\begin{equation}}
\def\eeq{\end{equation}}
\usepackage[all,2cell,dvips]{xy}

\def\IR{\relax{\rm I\kern -.18em R}}

\begin{document}
\title{The Hamiltonian structure of a coupled system derived from a supersymmetric breaking of Super KdV equations }
\author{ \Large  A. Restuccia*, A. Sotomayor**}
\maketitle{\centerline{*Departamento de F\'{\i}sica}}
\maketitle{\centerline{Universidad de Antofagasta }
\maketitle{\centerline{*Departamento de F\'{\i}sica}}
\maketitle{\centerline{Universidad Sim\'on Bol\'{\i}var}

\maketitle{\centerline{**Departamento de Matem\'aticas}
\maketitle{\centerline{Universidad de Antofagasta}}
\maketitle{\centerline{e-mail:arestu@usb.ve, asotomayor@uantof.cl
}}
\begin{abstract} A supersymmetric
breaking procedure for $N=1$ Super KdV, using a Clifford algebra,
is implemented. Dirac's method for the determination of
constraints is used to obtain the Hamiltonian structure, via a
Lagrangian, for the resulting solitonic system of coupled
Korteweg-de Vries type system. It is shown that the Hamiltonian
obtained by this procedure is bounded from below and in that
sense represents a model which is physically admissible.
\end{abstract}

Pacs: 12.60.Jv, 02.30.lk, 11.30.-j, 05.45. Yv, 02.30. Jr

\section{Introduction}
The coupled systems which are extensions of the Korteweg-de Vries
(KdV) equation arise in several physical problems and have
interesting properties. Such is the case of the coupled Ito
system \cite{Ito}, which describes the interaction of two
internal long waves. Among the properties of this system are the
existence of multisolitonic solutions which may be obtained using
the bilinear Hirota method \cite{Hirota,Yang}, the symmetries and
conserved quantities \cite{Mogorosi}, as well as the existence of
the Painlev\'e property, Lax pair and B\"{a}cklund transformations
\cite{Deng}.
 The supersymmetric KdV (SKdV) \cite{Mathieu1,Mathieu2} system is
described by coupled systems of partial differential equations in
terms of bosonic and fermionic fields. Such extensions of KdV
equation have infinite local and non-local conserved quantities
\cite{Gardner,Mathieu1,Mathieu2,Mathieu3,Dargis,Andrea1,Andrea2},
Lax pairs and at least one hamiltonian structure. Recently, using
a bosonization approach \cite{Andrea3}, exact solutions for the
$N=1$ (SKdV) \cite{Gao1,Gao2} and for the supersymmetric Ito
equation \cite{Ren} have been obtained. The bosonization approach
avoids to deal directly with Grassmann valued fields, not
suitable for many practical purposes, for example, in the search
for new solutions and in the analysis of the stability of
solitonic solutions.

An important aspect related to those systems is the supersymmetry
breaking and the resulting coupled equations derived from that
procedure. We consider here a supersymmetric breaking implemented
by replacing the Grassmann algebra by a Clifford algebra. This
scheme was already used in several works, see for example
\cite{Seibert}.

In that sense we present in this paper a coupled Korteweg-de Vries
system, with fields valued on a Clifford algebra, which has four
local conserved quantities and solitonic solutions, and obtain
its hamiltonian structure, with the consequent Poisson brackets
between the respective fields. For any Lagrangian system one can
always apply the Dirac's method for analyzing the constraints of
the theory and obtain from it the hamiltonian structure of the
system \cite{Dirac}. This approach was used in
\cite{Nutku,Kentwell} to obtain the, previously known, first and
second hamiltonian structures of KdV equation.

We also use the Dirac's method to obtain the hamiltonian
structure of the coupled system obtained from the supersymmetric
breaking scheme. We start obtaining a Lagrangian formulation for
the system then derive the hamiltonian via a Legendre
transformation and apply Dirac's method to obtain the constraints
in the phase space. It turns out that they are second class
constraints. We thus use the Dirac's brackets to obtain the
Poisson structure of the system in the constrained phase space. We
finally prove that the emerging hamiltonian is bounded from below
and this property grants a physically admissible content.

\section{Supersymmetry breaking procedure}
We denote by $u(x,t)$ and $\xi(x,t)$ the fields describing $N=1$
SKdV equations \cite{Mathieu1}, taking values at the even and odd
part of a Grassmann algebra respectively. The SKdV are

\beq\begin{array}{l}u_t=-u^{\prime\prime\prime}+6uu^\prime-3\xi\xi^{\prime\prime}
\\ \xi_t=-\xi^{\prime\prime\prime}+3{(\xi u)}^\prime.\label{eq1}
\end{array}\eeq
This system of partial differential equations have infinite local
conserved charges as well as infinite non-local conserved charges
\cite{Dargis,Andrea1,Andrea2}. The first few of them are
\beq\begin{array}{llll}H_{\frac{1}{2}}=\int_{-\infty}^{+\infty}\xi
dx,\\ H_1=\int_{-\infty}^{+\infty} u dx,
\\ H_3=\int_{-\infty}^{+\infty}\left(u^2-\xi\xi^\prime\right)dx,\\ H_5=\int_{-\infty}^{+\infty}
\left(2u^3+{(u^\prime)}^2-\xi^\prime\xi^{\prime\prime}-4u\xi\xi^\prime\right)dx,
\end{array}\label{eq2}\eeq
which we give explicitly in order to compare with the conserved
charges of the system with broken supersymmetry we will consider.

$H_1$ and $H_3$ become manifestly positive self-adjoint operators
in the quantum formulation of the theory. The fields $u$ and
$\xi$ may be expanded in terms of the Virasoro generators of a
superconformal algebra which may be realized in terms of an
oscillator algebra \cite{Mathieu2,Gervais,Nam}. Using normal
ordering the expressions of $H_1$ and $H$ are manifestly
positive. In particular $H_1$ is the hamiltonian of a
supersymmetric harmonic oscillator. Besides $H_1$ and $H_3$ we are
also interested in $H_{\frac{1}{2}}$ and $H_5$. $H_{\frac{1}{2}}$
will be relevant in a stability analysis and $H_5$ is very
important because stationary points of $H_5$ subject to $H_3$
give rise to solitonic solutions. Hence we would like to break
supersymmetry in a way to have a positive $H_3$ and conserved
charges analogous to $H_{\frac{1}{2}},H_1,H_5.$ It turns out that
$H_5$ is the hamiltonian of the new system obtained by breaking
supersymmetry.

To break supersymmetry we consider the fields $u$ and $\xi$ to
take values on a Clifford algebra instead of being Grassmann
algebra valued. We thus take $u$ to be a real valued field while
$\xi$ to be an expansion in terms of an odd number of the
generators $e_i,i=1,\ldots$ of the Clifford algebra:
\beq\xi=\sum_{i=1}^\infty
\varphi_ie_i+\sum_{ijk}\varphi_{ijk}e_ie_je_k+\cdots\label{eq3}\eeq
where \beq e_ie_j+e_je_i=-2\delta_{ij}\label{eq4}\eeq and
$\varphi_i,\varphi_{ijk},\ldots$ are real valued functions. We
define $ \bar{\xi}=\sum_{i=1}^\infty
\varphi_i\bar{e_i}+\sum_{ijk}\varphi_{ijk}\bar{e_k}\bar{e_j}\bar{e_i}+\cdots$
where $\bar{e_i}=-e_i$. We denote as in superfield notation the
body of the expansion those terms associated with the identity
generator and the soul the remaining ones. Consequently the body
of $ \xi\bar{\xi}$, denoted by $ \mathcal{P}(\xi\bar{\xi})$, is
equal to $\Sigma_i\varphi_i^2+\Sigma_{ijk}\varphi_{ijk}^2+\cdots$
In what follows, without lost of generality, we rewrite
$\sum_{i=1}^\infty \varphi_i^2+\sum_{ijk}\varphi_{ijk}^2$ simply
as $\mathcal{P}(\xi\bar{\xi})=\Sigma_i\varphi_i^2$.

We now propose the following system of partial differential
equations which has the required properties as discussed before,

\beq\begin{array}{ll}u_t=-u^{\prime\prime\prime}-uu^\prime-\frac{1}{4}{(\mathcal{P}(\xi\bar{\xi}))}^\prime
\\ \xi_t=-\xi^{\prime\prime\prime}-\frac{1}{2}{(\xi u)}^\prime.
\end{array}\label{eq5}\eeq

This system with a change of sign in the third term of the right
hand member of the first equation of system (5) was introduced
from a different point of view in \cite{Olver,Sokolov}. On the
other hand the system (5) in the particular case of only two
components is included in the classification given in
\cite{Foursov}.

The system (5) has the following conserved charges

\beq\begin{array}{llll}\hat{H_{\frac{1}{2}}}=\int_{-\infty}^{+\infty}\xi
dx,\\ \hat{H_1}=\int_{-\infty}^{+\infty} u dx,
\\ V\equiv \hat{H_3}=\int_{-\infty}^{+\infty}\left(u^2+\mathcal{P}(\xi\bar{\xi})\right)dx,
\\ M\equiv \hat{H_5}=\int_{-\infty}^{+\infty}
\left(-\frac{1}{3}u^3-\frac{1}{2}u\mathcal{P}(\xi\bar{\xi})+{(u^\prime)}^2+\mathcal{P}(\xi^\prime\bar{\xi^\prime})\right)dx.
\end{array}\label{eq6}\eeq

It is interesting to remark that the following non-local conserved
charge of Super KdV \cite{Andrea2} is also a non-local conserved
charge for the system (5), in terms of the Clifford algebra
valued field $\xi$,
\[\int_{-\infty}^\infty \xi(x)\int_{-\infty}^x\xi(s)dsdx.\]
 However the non-local conserved charges of Super KdV in \cite{Dargis} are not conserved by the system (5). For example, \[\int_{-\infty}^\infty
u(x)\int_{-\infty}^x\xi(s)dsdx\] is not conserved by (5).

We notice that \beq
\int_{-\infty}^{\infty}\mathcal{P}(\xi\bar{\xi})dx=\sum_{i=1}^\infty
\varphi_i^2=\|\xi\|^2_{L_2}\label{eq7}\eeq hence $V$ is manifestly
positive definite, one of the properties we were looking forward
th have. The system has solitonic solutions, for example:
$u(x,t)\equiv \phi(x,t)=3\mathcal{C}\frac{1}{\cosh^2(z)}, z\equiv
\frac{1}{2}\mathcal{C}^{\frac{1}{2}}(x-(1+\mathcal{C})t+a)$
($a\in \mathbb{R}$) , where  $\phi(x,t)$ is the one-soliton
solution of KdV equation, together with $\xi(x,t)=0$ is a one
soliton solution of the new system. In the same way the
multi-solitonic solutions of KdV together with $\xi(x,t)=0$ are
solutions of the new system.

The system (5) is not invariant under supersymmetric
transformations, as expected. Moreover, there isn't a conserved
charge of dimension 7, that is there is no analogue of $H_7$ as in
SKdV or KdV systems. The mechanism has not only broken
supersymmetry but also the symmetries related to $H_7$ and
probably to all higher local higher dimensional ones. There
remain, however, (6) as conserved charges of the system.

In the following section we will consider an extension of the
system (5) which depends on a parameter $\lambda$. For
$\lambda=1$ we will recover system (5). We will prove that the
system is Lagrangian and we will derive the Hamiltonian structure
of it.

\section{The Hamiltonian structure } We introduce the fields $w$
and $\eta_i$ defined by \beq u=w^\prime \mathrm{\:and\:}
\varphi_i=\eta_i^\prime\label{eq8}.\eeq

The following Lagrangian formulated in terms of $w$ and $\eta_i$
\beq S(w,\eta_i)\equiv \int dxdt\left[\frac{1}{2}w^\prime
\partial_tw+\frac{1}{6}{\left(w^\prime\right)}^3-\frac{1}{2}{\left(w^{\prime\prime}\right)}^2+\frac{1}{4}\lambda w^\prime
{\left(\eta_i^\prime\right)}^2+\frac{1}{2}\eta_i^\prime\partial_t\eta_i-\frac{1}{2}{\left(\eta_i^{\prime\prime}\right)}^2
\right],\label{eq9}\eeq where a repeated index $i$ implies
summation on that index, yields under variations of $w$ and
$\eta_i$ the system of equations

\beq\begin{array}{ll}\partial_tu=-u^{\prime\prime\prime}-\frac{1}{2}{\left(u^2\right)}^\prime-\frac{1}{4}\lambda{\left(\varphi_i^2\right)}^\prime,
\\ \partial_t\varphi_i=-\varphi_i^{\prime\prime\prime}-\frac{1}{2}\lambda{(u\varphi_i)}^\prime.
\end{array}\label{eq10}\eeq When $\lambda=1$ it reduces to (5),
the only case for which the dynamical equation for the field
$\xi$ is the same as the equation for the odd Grassmann field in
the Super KdV equations. It has the following local conserved
charges

\beq\begin{array}{llll}\tilde{H_{\frac{1}{2}}}=\int_{-\infty}^{+\infty}\xi
dx,\\ \tilde{H_1}=\int_{-\infty}^{+\infty} u dx,
\\ \tilde{H_3}=\int_{-\infty}^{+\infty}\left(u^2+\mathcal{P}(\xi\bar{\xi})\right)dx,
\\ \tilde{H_5}=\int_{-\infty}^{+\infty}
\left(-\frac{1}{3}u^3-\frac{\lambda}{2}u\mathcal{P}(\xi\bar{\xi})+{(u^\prime)}^2+\mathcal{P}(\xi^\prime\bar{\xi^\prime})\right)dx.
\end{array}\label{eq11}\eeq as well as the non-local conserved
charge \beq\int_{-\infty}^\infty
\varphi_i(x)\int_{-\infty}^x\varphi_i(s)dsdx.\label{eq12}\eeq We
will now show that $\tilde{H_5}$ is the Hamiltonian of the system
(10). We also remark that $\tilde{H_3}$ is equal to the $L_2$
norm of the Cifford algebra valued fields, \beq
\tilde{H_3}=\|(u,\xi)\|^2_{L_2}\geq0. \label{eq13}\eeq In order
to construct the Hamiltonian of system (10) we introduce the
conjugate momenta to $(w,\eta_i)$. They will be denoted by
$(p,\sigma_i)$.

We have \beq \begin{array}{lll}p:=\frac{\partial
\mathcal{L}}{\partial
(\partial_tw)}=\frac{1}{2}w^\prime=\frac{1}{2}u \\ \\
\sigma_i:=\frac{\partial \mathcal{L}}{\partial (\partial_t
\eta_i)}=\frac{1}{2}\eta_i^\prime=\frac{1}{2}\varphi_i.
\end{array} \label{eq14}\eeq Since from (14) we cannot obtain
$\partial_tw$ and $\partial_t\eta_i$ in terms of their conjugate
momenta, then these equations are primary constraints on the phase
space \cite{Dirac}.

In order to obtain the Hamiltonian of the system we perform a
Legendre transformation \beq
H=<p\partial_tw+\sigma_i\partial_t\eta_i-\mathcal{L}>_x\label{eq15}\eeq
where $<>_x$ denotes integration on the whole real line using $x$
as the integration variable.

We obtain \beq H=\frac{1}{2}\tilde{H_5},\label{eq16}\eeq where $
\tilde{H_5}$ is given in equation (11).

Following the Dirac approach to obtain the Hamiltonian structure
of equations (10), we consider the canonical Hamiltonian \beq
\tilde{H}=H+\left\langle\Lambda\left(p-\frac{1}{2}w^\prime\right)+\Lambda_i\left(\sigma_i-\frac{1}{2}\eta_i^\prime\right)\right\rangle_x,
\label{eq17}\eeq where $\Lambda$ and $\Lambda_i$ are Lagrange
multipliers.

The conservation of $p-\frac{1}{2}w^\prime$ yields \beq
\left\{p-\frac{1}{2}w^\prime,\tilde{H}\right\}=-w^\prime
w^{\prime\prime}-w^{\prime\prime\prime\prime}-\frac{\lambda}{4}\Lambda{\left({\left(\eta_i^\prime\right)}^2\right)}^\prime-\Lambda^\prime=0
\label{eq18}\eeq and the conservation of
$\sigma_i-\frac{1}{2}\eta_i^\prime=0$ implies \beq
\left\{\sigma_i-\frac{1}{2}\eta_i^\prime,\tilde{H}\right\}=-\frac{\lambda}{2}{\left(w^\prime\eta_i^\prime\right)}^\prime-\eta_i^{\prime\prime\prime\prime}
-\Lambda_i^\prime=0, \label{eq19}\eeq where $\left\{,\right\}$
denotes the Poisson bracket on the original unconstrained phase
space.

 (18) and (19) determine the Lagrange multipliers $\Lambda$
and $\Lambda_i$ respectively. Hence the Dirac procedure ends up
with these equations. There are no more contraints in the phase
space.

It turns out that both constraints (14) are second class ones.

In fact, if we denote

\beq\begin{array}{ll}v:=p-\frac{1}{2}w^\prime \\
v_i:=\sigma_i-\frac{1}{2}\eta_i^\prime
\end{array} \label{eq20}\eeq and $v_I:=(v,v_i)$, we then have
\beq\left\{v_I(x),v_J(x^\prime)\right\}=-\delta_{IJ}\partial_x\delta(x-x^\prime).\label{eq21}\eeq

The Poisson structure of the constrained Hamiltonian is then
determined by the Dirac brackets \cite{Dirac}. For any two
functionals on the phase space $F$ and $G$, the Dirac bracket is
defined as
\beq\left\{F,G\right\}_{DB}:=\left\{F,G\right\}-\langle\langle\left\{F,v_I(x^\prime)\right\}
\left\{v_I(x^\prime),v_J(x^{\prime\prime})\right\}^{-1}\cdot\left\{v_J(x^{\prime\prime}),G\right\}
\rangle_{x^\prime}\rangle_{x^{\prime\prime}},\label{eq22}\eeq
where
\beq\langle\left\{v_I(x^\prime),v_J(x^{\prime\prime})\right\}^{-1}g(x^{\prime\prime})
\rangle_{x^{\prime\prime}}=-\delta_{IJ}\int_{-\infty}^{x^\prime}g(\tilde{x})d\tilde{x}.\label{eq23}\eeq

We then have
\beq\begin{array}{lll}\left\{u(x),u(y)\right\}_{DB}=\partial_x\delta(x,y),\\
\left\{\varphi_i(x),\varphi_j(y)\right\}_{DB}=\delta_{ij}\partial_x\delta(x,y),\\\left\{u(x),\varphi_i(y)\right\}_
{DB}=0.\end{array}\label{eq24}\eeq Consequently,
\beq\begin{array}{ll}\partial_tu=\left\{u,H\right\}_{DB}=-\frac{1}{2}{(u^2)}^\prime-u^{\prime\prime\prime}-
\frac{\lambda}{4}{(\varphi_i^2)}^\prime\\
\partial_t\varphi_i=\left\{\varphi_i,H\right\}_{DB}=-\varphi_i^{\prime\prime\prime}-\frac{\lambda}{2}{(u\varphi_i)}^\prime,
\end{array}\label{eq25}\eeq where $H$ is given by (16) and can be directly expressed in
terms of $u$ and $\xi$. Notice that
$\left\{u,\tilde{H}\right\}_{DB}=\left\{u,H\right\}_{DB}$ and the
same happens for any function of $u$ and $\varphi_i$.

We have then derived the Poisson structure of the Hamiltonian
system. It follows directly from the existence of a Lagrangian
for the dynamical system. The Dirac procedure determines the
constraints of the phase space together with the Poisson
structure. The final Poisson structure is obtained from the Dirac
bracket (which satisfies the properties of a Poisson bracket, in
particular the Jacobi identity).

\section{Properties of the Hamiltonian}
An important property of the Hamiltonian of a physical system is
that it is bounded from below. The Hamiltonian of the Super KdV
system satisfies that property, in fact it is positive. We now
show that the Hamiltonian $H$ of the dynamical system arising
from the breaking of supersymmetry indeed has also this property.

We consider
\[\tilde{H_3}+\tilde{H_5}=\|(u,\xi)\|^2_{H_1}+\int_{-\infty}^{+\infty}\left(-\frac{1}{3}u^3-\frac{\lambda}{2}u\mathcal{P}(\xi\bar{\xi})\right)dx\]
where the Sobolev norm $\|\|_{H_1}$ is defined by
\[\|(u,\xi)\|^2_{H_1}:=\int_{-\infty}^{+\infty}\left[u^2+\mathcal{P}( \xi\bar{\xi})+{u^\prime}^2+\mathcal{P}( \xi^\prime\bar{\xi^\prime})\right]dx.\]
We also noticed that
\[\tilde{H_3}=\|(u,\xi)\|^2_{L^2}\] where $\|\|_{L^2}$ is the $L^2$
norm.

We then have
\[\tilde{H_3}+\tilde{H_5}\geq \|(u,\xi)\|^2_{H_1}-\frac{m}{2}\int_{-\infty}^{+\infty}|u|\left(u^2+\mathcal{P}(\xi\bar{\xi}\right)dx \]
where $m=\max(1,|\lambda|).$

We now use the bound

\[\sup|u|\leq \frac{\|u\|_{H_1}}{\sqrt{2}}\leq
\frac{\|(u,\xi)\|_{H_1}}{\sqrt{2}},\]to  obtain
\[\tilde{H_3}+\tilde{H_5}\geq \|(u,\xi)\|^2_{H_1}-\frac{m}{2\sqrt{2}}\|(u,\xi)\|_{H_1}\|(u,\xi)\|_{L^2}.\]

Consequently
\[\tilde{H_3}+\tilde{H_5}+{\left(\frac{m}{4\sqrt{2}}\right)}^2\tilde{H_3}\geq {\left(\|(u,\xi)\|_{H_1}-\frac{m}{4\sqrt{2}}\|(u,\xi)\|_{L^2}\right)}^2\geq0.\]
Finally
\[\tilde{H_5}\geq-\left(1+{\left(\frac{m}{4\sqrt{2}}\right)}^2\right)\tilde{H_3},\]
hence for a normalized state satisfying $\|(u,\xi)\|_{L^2}=1$ we
have
\[\tilde{H_5}\geq-\left(1+{\left(\frac{m}{4\sqrt{2}}\right)}^2\right).\]
The Hamiltonian is then bounded from below in the space of
normalized $L_2$ configurations and it is thus physically
admissible.

We notice that is important to have
$\tilde{H_3}=\|(u,\xi)\|^2_{L^2}$ in order to conclude this
property of the Hamiltonian. The dynamical system (10) for any
value of $\lambda$ has then a Hamiltonian structure which is
physically admissible.

For the particular value $\lambda=2$, the system (10) is
invariant under the Galileo transformations
\[\begin{array}{llll}t\rightarrow \hat{t}=t \\ x\rightarrow \hat{x}=x+ct \\ u\rightarrow \hat{u}=u+c \\ \varphi_i\rightarrow
\hat{\varphi_i}=\varphi_i.\end{array}\] However for this
particular value of $\lambda$ the system decouples into
independent KdV equations.

The dynamical system (10) has multi-solitonic solutions
corresponding to $u$ a multi-soliton of KdV and $\xi=0$. It is
well known that this solutions are Liapunov stable for the KdV
equation \cite{Benjamin,Bona}. The same stability problem for the
system (10) cannot be deduced straightforwardly from it since
small initial perturbations of $\xi$ may grow to produce
instabilities in the system. A detailed analysis of this problem
is presented in \cite{Alvaro}, where we show that the solitonic
solutions of the system (5) has nice stability properties. On the
other side, it is straightforward to show that the solution
$u=\xi=0$ corresponding to the minimum of $\tilde{H_3}$ is $L^2$
stable. In fact, given $\epsilon$, since
$\tilde{H_3}=\|(u,\xi)\|_{L^2}$ is conserved under the dynamics of
system (10), if initially the perturbation satisfies
$\|(u,\xi)\|_{L^2}<\delta<\epsilon$ then is also bounded by
$\epsilon$ for any $t>0$.

\section{Conclusions} We discussed a solitonic KdV coupled system
defined in terms of a Clifford algebra. It was derived from a
supersymmetric breaking of $N=1$ SKdV equation. This procedure
not only breaks the supersymmetry but also the symmetries related
to the higher order local conserved quantities of Super KdV
equations. Only a finite number of local conserved quantities
remain valid. Nevertheless these conserved quantities are enough
to ensure nice stability properties of the solitonic solutions
\cite{Alvaro}. The coupled system has also non-local conserved
quantities, a non trivial one is explicitly shown in the paper.
It is exactly the same conserved quantity that appears in the
SKdV $N=1$ system.

We obtained the hamiltonian structure and the consequent Poisson
bracket of the system, using the Dirac's method for analyzing the
constraints. We also proved that the emerging hamiltonian is
bounded from below and in that sense it has a physically
admissible content.

We believe that the procedure followed in this work can be used
in several of the known KdV coupled systems.

$\bigskip$

\textbf{Acknowledgments}

A. R. and A. S. are partially supported by Project Fondecyt
1121103, Chile.


\begin{thebibliography}{}
\bibitem{Ito}{M. Ito, Phys. Lett. A 91, 335-338
(1982).}
\bibitem{Hirota}{R. Hirota and J. Satsuma, Phys. Lett. A 85,
407-408 (1981).}
\bibitem{Yang}{Yang Jian-Rong and Mao Jie-Jian, Commun. Theor.
Phys 49, 22-26 (2008).}
\bibitem{Mogorosi}{E. T. Mogorosi, B. Muatjetjeja and C. M.
Khalique, Boundary Value Problems 2012, 2012:150.}
\bibitem{Deng}{Deng Shan Wang, Appli. Math. Comp. 216, 1349-1354 (2010).}
\bibitem{Mathieu1}{P. Mathieu, J. Math. Phys. 29, 2499
(1988).}
\bibitem{Mathieu2}{P. Labelle and P. Mathieu, J. Math. Phys. 32,
923-927 (1990).}
\bibitem{Gardner}{R. M. Miura, C. S. Gardner, and M. D. Kruskal, J. Math. Phys. 9,
1204 (1968).}
\bibitem{Mathieu3}{P. Mathieu, Phys. Lett. B 203, 287-291
(1988).}
\bibitem{Dargis}{P. Dargis and P. Mathieu, Phys. Lett. A 176,
67-74 (1993).}
\bibitem{Andrea1}{S. Andrea, A. Restuccia and A. Sotomayor, J. Math.
Phys. 46, 103517 (2005).}
\bibitem{Andrea2}{S. Andrea, A. Restuccia and A. Sotomayor, Phys. Lett.
A 376, 245-251 (2012).}
\bibitem{Andrea3}{S. Andrea, A. Restuccia and A. Sotomayor, J. Math.
Phys.42, 2625 (2001).}
\bibitem{Gao1}{X. N. Gao and S. Y. Lou, Phys. Lett. B 707, 209 (2012).}
\bibitem{Gao2}{X. N. Gao, S. Y. Lou and X. Y. Tang, JHEP 05, 029 (2013).}
\bibitem{Ren}{B. Ren, J. Lin and J. Yu, Aip Advances 3 , 042129/12 (2013).}
\bibitem{Seibert}{N. Seibert, JHEP 0306, 010
(2003).}
\bibitem{Dirac}{P. A. M. Dirac, ``Lectures on Quantum Mechanics",
Belfer Graduate School Monograph Series No.2, Yeshiva University,
New York, 1964.}
\bibitem{Nutku}{Y. Nutku, J. Math. Phys. 25 (6) , June (1984).}
\bibitem{Kentwell}{G. W. Kentwell, J. Math. Phys. 29, 46 (1988).}
\bibitem{Gervais}{J. L. Gervais and A. Neveau, Nucl. Phys. B 209,
125 (1982).}
\bibitem{Nam}{S. Nam, Phys. Lett. B 172, 323
(1986).}
\bibitem{Benjamin}{T. B. Benjamin, Proc. R. Soc. Lond. A. 328,
153-183 (1972).}
\bibitem{Bona}{J. Bona, Proc. R. Soc. Lond. A.  344, 363-374
(1975).}
\bibitem{Olver}{P. J. Olver and V. V. Sokolov, Commun. Math. Phys. 193(2),
245-268 (1998).}
\bibitem{Sokolov}{V. V. Sokolov and S. I. Svinolupov, Theor. Math. Phys. 100 B,
959-962 (1994).}
\bibitem{Foursov}{M. V. Foursov,  J. Math. Phys. 44, 3088, (2003).}
\bibitem{Alvaro}{A. Restuccia and A. Sotomayor, J. Phys.: Conf. Ser. 410 012073 (2013) .}


\end{thebibliography}
\end{document}